\documentclass{article}

\usepackage{arxiv}

\usepackage[utf8]{inputenc} 
\usepackage[T1]{fontenc}    
\usepackage{hyperref}       
\usepackage{url}            
\usepackage{booktabs}       
\usepackage{amsfonts}       
\usepackage{nicefrac}       
\usepackage{microtype}      
\usepackage{lipsum}

\usepackage{listings}
\usepackage[table,xcdraw]{xcolor}

\definecolor{codegreen}{rgb}{0,0.6,0}
\definecolor{codegray}{rgb}{0.5,0.5,0.5}
\definecolor{codepurple}{rgb}{0.58,0,0.82}
\definecolor{backcolour}{rgb}{0.95,0.95,0.92}

\lstdefinestyle{mystyle}{
    backgroundcolor=\color{backcolour},   
    commentstyle=\color{codegreen},
    keywordstyle=\color{magenta},
    numberstyle=\tiny\color{codegray},
    stringstyle=\color{codepurple},
    basicstyle=\ttfamily\footnotesize,
    breakatwhitespace=false,         
    breaklines=true,                 
    captionpos=b,                    
    keepspaces=true,                 
    numbers=left,                    
    numbersep=5pt,                  
    showspaces=false,                
    showstringspaces=false,
    showtabs=false,                  
    tabsize=2
}

\usepackage{graphicx,graphics}

\title{On the benchmarking of partitioned \\real-time systems}

\author{
  Felipe Gohring de Magalhaes, Alexy Torres Aurora Dugo, \\ 
    \textbf{Jean-Baptiste Lefoul and Gabriela Nicolescu} \\
    Department of Computer Engineering and Software Engineering\\
    Ecole Polytechnique de Montreal \\
    Montreal, QC, Canada \\  
    \texttt{contact: felipe.gohring-de-magalhaes@polymtl.ca} \\
}

\begin{document}
\maketitle

\begin{abstract}
    Avionic software is the subject of critical real time, determinism and safety constraints. Software designers face several challenges, one of them being the estimation of worst-case execution time (WCET) of applications, that dictates the execution time of the system. A pessimistic WCET estimation can lead to low execution performances of the system, while an over-optimistic estimation can lead to deadline misses, breaking one the basic constraints of critical real-time systems (RTS). Partitioned systems are one special category of real time systems, employed by the avionic community to deploy avionic software. The ARINC-653 standard is one common avionic standard that employs the concept of partitions. This standard defines partitioned architectures where one partition should never directly interfere with another one. Assessing WCET of general purpose RTSs is achievable by the usage of one of the many published benchmark or WCET estimation frameworks. Contrarily, partitioned RTSs are special cases, in which common benchmark tools may not capture all the metrics. In this document, we present SFPBench, a generic benchmark framework for the assessment of performance metrics on partitioned RTSs. The general organization of the framework and its applications are illustrated, as well as an use-case, employing SFPBench on an industrial partitioned operating system (OS) executing on a Commercial Off-The-shelf (COTS) processor. 
 
\end{abstract}


\section{Introduction}
    Avionic applications must follow strict certification rules. They are hard real-time systems, which means that they are considered to have failed if they do not completer their functions within stringent deadlines. Predictability and determinism of such systems is crucial and are major components of certification for aerospace systems. This is due to the fact that these systems are one of the most common critical systems which civilians have daily access to. What differentiates avionic systems to others is their catastrophe potential. While a failing ship can easily stop and wait for maintenance, as much as a problematic car can park on the side of the road, where usually there is low to no risk involved, a failing airplane cannot simply stop in the middle of a trip for repairs.

    Operating Systems (OSs) provide an interface between the software applications and their host hardware systems. One of their main roles is to  manage the system's resources to make their usage transparent to the user. This enables applications to run on hardware systems without being aware of these systems' specifications, easing the development of applications. Real-Time Operating Systems (RTOS) are operating systems designed for real-time management. RTOSs are used for critical systems for which a given functionality must be done within a given time interval. An RTOS must ensure that the worst-case execution time (WCET) of each task is respected. The ARINC-653 \cite{arinc653} is an avionic standard which defines a general-purpose APplication/EXecutive (APEX) interface (API) between the Core Software (CSW) of an Avionics Computer Resource (ACR) and the application software. ARINC-653 compliant RTOS extend the functionalities provided by regular RTOS enabling time and space isolation of applications.  
    
    Assessing RTOS performances is performed either using intrusive methods or non-intrusivy methods. Intrusive assessment is usually precise, but requires access to all modules of the RTOS (i.e. source code). Performance metrics are directly read during execution time relying on defined points in the RTOS modules. Non-intrusive assessment is used when not all the modules are available (i.e. only the APIs are available, but not their implementation). In this scheme, the performance metrics are obtained using only API calls, collecting the metrics before and after their execution. The main advantage of intrusive assessment is the seamless access to specific points of the OS, while non-intrusive assessment eases the portability of tests, enabling the generalization of tests. Assessing WCET estimation can be done statically or dynamically \cite{wcet}. For safety reasons, there is always a margin added to the estimated WCET. By providing a more precise WCET estimation, it is possible to reduce this margin, leading to more tasks being scheduled and improve the hardware resources usage. The determinism of execution time can be measured with its standard deviation. The smaller the standard deviation is, the more deterministic the system execution time is. The determinism is critical for real-time systems, since it can guarantee the timing behavior of a system. 

    State-of-the-art RTOSs (i.e. VxWorks \cite{vxworks}, PikeOS \cite{pikeos}, Integrity178 \cite{integrity}) are characterized by performance metrics showing their efficiency using in-house assessments. Although published results show precise metrics, it is a fairly difficult task for third-part users / developers to replicate these results in the context of their applications. To tackle this limitation, different solutions seek on providing standard benchmark frameworks, replicable by any provider or user. These enable a common point for different providers to assess their RTOS performances as well as compare with competitors. Nevertheless, ARINC-653 compliant RTOSs cannot profit from these benchmark frameworks, due to the special characteristics of ARINC-653 compliant RTOSs. ARINC-653 systems have time and space division and follow specific design recommendations, such flushing the entire cache memory for each partition switch , which differs them from general RTOSs.
    
    This document presents an open-source benchmark framework for ARINC-653 compliant RTOSs, called \textbf{S}traight-\textbf{F}orward \textbf{P}artitions-Aware \textbf{Bench}mark framework (SFPBench). Different classes of benchmark applications are developed in order to assess different metrics of the RTOS. The tests are organized as gray-box applications relying solely on standard ARINC-653 APEX interface. The benchmark seeks to be openly accessible and modifiable by the community, hosted on the public domain (i.e. github) \cite{github_sff}.
    
    The remainder of this document is organized as it follows. Next section presents an overview of the ARINC-653 standard, followed by Section \ref{sec:sfpbench}, where the general organization of SFPBench and its applications is introduced. Section \ref{sec:p2020} illustrates the validation hardware structure as well as the system configuration parameters. Next, Section \ref{sec:results} presents the results obtained using SFPBench with an industrial RTOS. Finally, \ref{sec:conclusion} concludes this documents and points to possible extentions of SFPBench functionalities.

\section{Avionics systems and the ARINC-653 standard}
\label{sec:arinc}
    Former avionic systems relied on stand-alone architectures to deploy applications. The main motivation was due to security reasons, in a way to isolate each one of the components of the system on its own domain, without any interference between applications. In the beginning of the twenty-first century, avionics systems underwent the transition from federated, isolated, architectures to Integrated Modular Avionic (IMA) architectures \cite{4391842}. The motivation was to reduce Size, Weight, and Power (SWaP) issues, which are common to most embedded systems.

\begin{figure}[!htb]
      \centering
      \includegraphics[width=15cm]{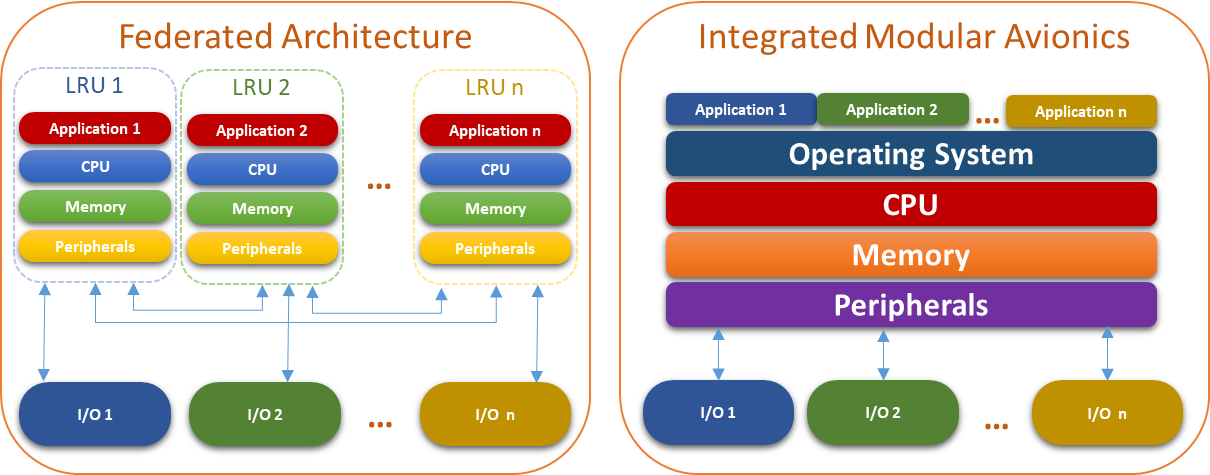}
      \caption{Federated and IMA architectures comparison.}
      \label{fig:IMA_federated}
    \end{figure}

    Figure \ref{fig:IMA_federated} illustrates the difference between applications deployed with federated architectures and IMA. In a federated architecture, each application has its own hardware called a Line Replaceable Unit (LRU). LRUs can be seen as a set of interconnected boxes. The drawback of these easily replaceable hardware units is the cost of redundancy of hardware. This is one of the reasons IMA architectures are used today: one computing unit can be used to support multiple applications, allowing a hardware computing unit to be used by multiple applications, therefore the cost of redundancy is reduced.
    
    One of the most common standards being used by the avionic society is the ARINC-653 \cite{arinc653} standard. It is a standard for partitioned RTOS which gives specifications to ensure the isolation of applications. According to this standard, the partitioning must be done in space and in time. 

    \paragraph{Space partitioning} Each partition is isolated regarding hardware usage, such as memory space: each partition has a set of addresses in memory and it is the only one having the rights to access them.

    \paragraph{Time partitioning} The CPU time is divided in several time windows. Each time window is allocated to a partition. During one of its time windows, a partition is the only one executing on a CPU. All partitions are allocated time windows within a period of time called major time frame. The schedule is then repeated every major time frame. 

    The standard also specifies services that the RTOS must offer. These services are called APEX services.

    \paragraph{APEX API} The services offered by this API are the ones used to create the ARINC-653 partitions and perform synchronization and communication between them. An ARINC-653 partition can be composed of several ARINC-653 processes that share the partition's context. An analogy with POSIX's API would be that ARINC-653 partitions are POSIX processes and ARINC-653 processes are POSIX threads.

    \paragraph{Interpartition communication} ARINC-653 specifies how two partitions can communicate. The communication means are messages using channels or ports. There are two modes of communication:
    \begin{itemize}
        \item[-] Sampling mode: only one message is stored in the source port, it is overwritten each time the source partition writes. It is useful when a partition requires the latest status of a data.
        \item[-] Queuing mode: messages sent are stored in a FIFO order. Each partition (sender and receiver) is responsible in handling the situations when the queue is full or empty.
    \end{itemize}

    \paragraph{Intrapartittion communication} There are several communication means between the processes of a same partition: 
    \begin{itemize}
        \item[-] Blackboard: very similar to sampling mode; instead of being between partitions, it is between ARINC-653 processes.
        \item[-] Buffers: very similar to queue mode; instead of being between partitions, it is     between ARINC-653 processes.
        \item[-] Semaphore: ARINC 653 semaphores conform with the classical definition:
        WAIT\_SEMAPHORE is called to wait on a semaphore (if the value of semaphore not equal to zero, the semaphore is decremented and the process continues, or else it is blocked until the semaphore is incremented). SIGNAL\_SEMAPHORE is used to increment the semaphore's value and potentially freeing a locked process. Waiting processes are queued in FIFO order, and freed one at a time. 
        \item[-] Events: processes can wait on custom events, which have two states ("up" if the event occurred or "down" if not). All processes waiting on an event with a "down" state are blocked until either they timed out or the event's state changes. When an event is "up" all waiting processes are freed at the same time, making all of them candidates to be scheduled, unlike for semaphores.
        \item[-] Mutex: as semaphores, ARINC-653 mutexes conform with the classical definition. A mutex can be owned by only one process at a time. Waiting processes are queued in a FIFO, similarly to semaphores.
    \end{itemize}

    \paragraph{Health Monitor} The Health Monitor is a feature of the RTOS which must handle unexpected error during the execution of partitions, such as deadline misses or arithmetical errors. Through configurations by the user, the Health Monitor then decides what behavior the partition must have, whether it must shutdown or reset the partition, ignore the error or try recovering from it. The ARINC-653 standard requires the RTOS to have a Health Monitor.
    
\section{SFPBench - Collaborative open source benchmark framework}
\label{sec:sfpbench}
    Benchmarking real-time systems is a widely studied subject, where several approaches are possible. Different solutions opt for to exploit specifics metrics \cite{bench1} or require specialized libraries or system modifications to be used \cite{bench2}\cite{bench3}. These design decisions affect the level of accuracy of the evaluations obtained by using each solution. Another important aspect is the type of access the end-user has regarding the target system implementation. Usually, the end-user, which its deploying its application on the target system, only has access to the RTOS services through APIs, masking the under-layers design aspects. This imposes several restrictions to test the system, as the abstraction level can be too elevated to actually capture desired metrics. Ideally, benchmark applications should require little or no modification in the host RTOS, for the collected results to be less prone to interferences. In \cite{bench4}, a benchmark framework is presented, where no modification in the host RTOS or the usage of specific libraries is demanded. The set of tests is meant for basic services, such as task switch time and semaphore latency. A set of benchmark applications is introduced in \cite{bench3}. These applications do not require additional libraries nor specific RTOS services, such as mutexes or pthreads, serving as good metrics to establish worts-case execution time of host systems. Although these may serve as good common baseline applications for systems' performance assessment, their contribution is limited by their conception. Most important, these solutions are not meant for ARINC-653 compliant RTOSs, jeopardizing their usage for the context of this document. 

\begin{figure}[!htb]
      \centering
      \includegraphics[width=7.5cm]{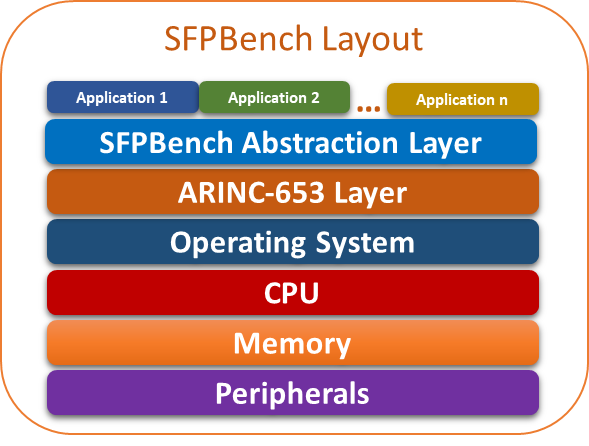}
      \caption{SFPBench organization layout, showing the abstraction layer between the ARINC-653 implementation layer and deployed application(s).}
      \label{fig:SFPBench}
    \end{figure}

    Respecting the basic premise of ARINC-653 systems, in which partitions are memory and time isolated, SFPBench framework is deployed to stress specific points of such systems. The benchmark framework provides several benchmark applications developed to assess different aspects of the RTOS, such as partition and process switch times, semaphore latency as well as covering ARINC-653 APEX calls latency. The framework is organized such as the applications do not require any modification to execute with the target RTOS, hence eliminating any source of interference in the application level. To enable that, an abstraction layer is provided as part of the framework, where RTOS specific definitions are linked. Run-time macros (i.e. \textit{START\_TIME\_MEASURE} and \textit{END\_TIME\_MEASURE}) are provided with the porting layer such as new applications and performance metrics can be easily added to the original benchmark set. The measurements rely on the hardware clock tick of the system, thus enhancing the measurement accuracy. Figure \ref{fig:SFPBench} illustrates the SFPBench abstraction layer placement in the ARINC-653 compliant system. As the figure shows, the abstraction layer is placed between the ARINC-653 RTOS and the deployed application. 
    
    The main purpose of this layer is to abstract the API calls used by the test application. This enables the application to remain completely agnostic of the host RTOS, making it possible to use SFPBench on different targets RTOSs without modifying the application. The only requirement is to update the abstraction layer, mapping the host RTOS API to the standard defined for SFPBench. 
    
    Listing \ref{list:example} presents an example of a function definition in the abstraction layer. First, a generic function to create processes is declared: \textit{perf\_create\_task}. As the target system is an ARINC-653 compliant RTOS, the function body implements (between lines 1 and 37)  the default process creation. The application main function (partition entry point) can simply call this function (\textit{perf\_create\_task}) and the process creation is performed in the abstraction layer. This is illustrated between lines 39 and 47.\newline 
  
\begin{lstlisting}[firstline=1,lastline=47, language=C, caption=Abstraction layer function example., label=list:example]
perf_task_handle_t perf_create_task(perf_task_entry_t task_entry,
                                    char task_name[4],
                                    unsigned int prio)
{
  perf_task_handle_t thandle;
  RETURN_CODE_TYPE errCode = NO_ERROR;

  PROCESS_ATTRIBUTE_TYPE processAttributes;
  processAttributes.BASE_PRIORITY = (PRIORITY_TYPE)prio;
  processAttributes.DEADLINE = SOFT;
  processAttributes.ENTRY_POINT = (SYSTEM_ADDRESS_TYPE)task_entry;
  strncpy(processAttributes.NAME, task_name, MAX_NAME_LENGTH);
  processAttributes.PERIOD = INFINITE_TIME_VALUE;
  processAttributes.STACK_SIZE = 1024;
  processAttributes.TIME_CAPACITY = INFINITE_TIME_VALUE;
  CREATE_PROCESS(&processAttributes, &thandle, &errCode);

  if(errCode != NO_ERROR)
  {
    PERF_PRINT_STRING("Just had an error creating the task:  ");
    PERF_PRINT_NUMBER(errCode);
    PERF_PRINT_EOL();
  }

  else
  {
    START(thandle, &errCode);
    if(errCode != NO_ERROR)
    {
      PERF_PRINT_STRING("Just had an error starting the task:  ");
      PERF_PRINT_NUMBER(errCode);
      PERF_PRINT_EOL();
    }
  }

  return thandle;
}
////
void main()
{
  RETURN_CODE_TYPE retCode = NO_ERROR;

  perf_create_task(RunTest, "Test1", BASE_PRIO);

  SET_PARTITION_MODE(NORMAL, &retCode);

}
\end{lstlisting}
    
    To collect execution metrics, four macros are defined in the abstraction layer:
    \begin{itemize}
        \item DECLARE\_TIME\_MEASURE()
        \item INITIALIZE\_TIME\_VARS(name)
        \item START\_TIME\_MEASURE()
        \item END\_TIME\_MEASURE()
        \item VALIDATE\_TIME\_MEASURE(std\_variation\_enable)
    \end{itemize}
    
    These macros can be used to collect performance metrics for any application or function, provided by the framework or not. Their usage is straight-forward: one should place the desired point of measurement between the two macros - START\_TIME\_MEASURE() and END\_TIME\_MEASURE(). Listing \ref{list:sobel} presents an usage example of these macros. In the example, the main() function creates a process called \textit{RunSobelTests} (line 4). The process body implements the execution of the test: control variables are declared (line 12); performance variables are declared and initialized for the "SOBEL" test (lines 13 and 14); the execution time recording is started before the target functions are called (line 22) and the measured functions are executed (lines 23 and 24); after execution of the target functions, the time collection is stopped and the results are calculated and validated (lines 25 and 26); after executing the same tests a given amount of times (configured in the for..loop on line 15), the test is finalized and the obtained results exhibited (line 28). Please notice that the implementation of tested functions (do\_gaussian and do\_sobel) and most of control variables are omitted for the sake of better readability.
    
\begin{lstlisting}[firstline=49,lastline=79, language=C, caption=Performance macros usage example., label=list:sobel]
void main()
{
  RETURN_CODE_TYPE retCode = NO_ERROR;
  perf_create_task(RunSobelTests, "SOBEL", BASE_PRIO);
  SET_PARTITION_MODE(NORMAL, &retCode);
  while(1)
}

void  RunSobelTests ()
{
  int32_t runs, j;
  DECLARE_TIME_MEASURE()
  INITIALIZE_TIME_VARS("SOBEL")
  for(runs = 0; runs < SOBEL_RUNS_PER_MEASURE; runs++)
    {
      for (j = 0; j < height * width; j++)
        {
          image[j] = (uint8_t) rand();
        }

      INIT_TIME_MEASURE()
      do_gaussian();
      do_sobel();
      FINISH_TIME_MEASURE()
      VALIDATE_TIME_MEASURE(1)
    }
  PRINT_PERFORMANCE_INFO()
  perf_task_suspend_self();
  while(1); 
}
\end{lstlisting}
    
    The SFPBench framework - at the state of the time of the submission of this document - provides eighteen (18) independent test applications, divided in three groups: grey-box applications, APEX API and complete applications. Each one is presented next.
    
    \subsection{Grey-box applications}
    \label{sub:grey}
        These test applications are developed with the intent of testing specific RTOS metrics, such as the general overhead to use a given service (i.e., semaphore) in the context of an application. This group of tests is composed by the following applications:
        \begin{itemize}
            \item \textbf{Process Switch} creates \textit{'n'} processes and collects the time between the end of a process and the begining of the next one. This enables the assessment of the process switch time of the target RTOS, where the RTOS overhead to switch between two processes can be evaluated;
            \item \textbf{Mutex Acquire and Mutex Release} creates different processes that use mutex primitives to define a critical region. The collected time refers to the time it takes for one process to request access to a mutex and for another process to release it;
            \item \textbf{Mutex Acquire 2 and Mutex Release 2} similarly to the previous test, this test creates different processes that use mutex primitives to define a a critical region. The difference is in the moment the time collection is enabled. In this test, the total time to loop the execution is collected;
            \item \textbf{Mutex Workload} creates two process that dispute a critical region. When in the the critical region, the process with access granted performs a series of operations within the region, emulating the processing of data in a real critical region;
            \item \textbf{Sem Wait and Sem Signal} creates different processes that use semaphores primitives to synchronize the execution of the application. The collected time refers to the time it takes for one process to request access to a wait on a semaphore and for another process to release it;
            \item \textbf{Priority Sem} creates processes that compete for a semaphore, each with a different priority;
            \item \textbf{Sem Signal 2 and Sem Wait 2} similarly to the previous test, this test creates different processes that use semaphores primitives to synchronize the execution of the application. The difference is in the moment the time collection is enabled. In this test, the total time to loop the execution is collected;
            \item \textbf{Sem Workload} creates two process that use semaphores primitives to synchronize the execution of the application. When a process passes the wait semaphore primitive, it performs a series of operations, emulating the processing of data, and;
            \item \textbf{Partition Switch} measures the time before the end of one partition and the beginning of the next. This enables the assessment of the partition switch time of the target RTOS, where the RTOS overhead to switch between two partitions can be evaluated.
        \end{itemize}
    
    \subsection{APEX API}
    \label{sub:apex}
        This group has one application that is designed to individually assess the performance of each one of the APEX calls defined in the ARINC-653 Part 1 documentation. The complete list of calls is presented in Table \ref{tab:apex}.
    
        \begin{table}[!htb]
            \scriptsize
            \centering
            \caption{Covered APEX Calls.}
            \begin{tabular}{c|c|c|c}
            \textit{GET\_PARTITION\_STATUS} & \textit{CREATE\_SEMAPHORE} & \textit{CREATE\_BUFFER} & \textit{CREATE\_BLACKBOARD}  \\ \hline
            \textit{READ\_BLACKBOARD} & \textit{GET\_BUFFER\_ID} & \textit{SEND\_BUFFER} & \textit{RECEIVE\_BUFFER}  \\ \hline
            \textit{DISPLAY\_BLACKBOARD} & \textit{WAIT\_SEMAPHORE} & \textit{SET\_PRIORITY} & \textit{GET\_MY\_ID}   \\ \hline
            \textit{GET\_SEMAPHORE\_STATUS} & \textit{CREATE\_EVENT} & \textit{SET\_EVENT} & \textit{GET\_EVENT\_ID}   \\ \hline
            \textit{GET\_CURRENT\_TICKS} & \textit{CREATE\_QUEUING\_PORT} & \textit{GET\_QUEUING\_PORT\_ID} & \textit{GET\_QUEUING\_PORT\_STATUS}   \\ \hline
            \textit{SEND\_QUEUING\_MESSAGE} & \textit{RECEIVE\_QUEUING\_MESSAGE} & \textit{WRITE\_SAMPLING\_MESSAGE} & \textit{READ\_SAMPLING\_MESSAGE}  \\ \hline
            \textit{SIGNAL\_SEMAPHORE} & \textit{GET\_PROCESS\_STATUS} & \textit{WAIT\_EVENT} & \textit{GET\_SAMPLING\_PORT\_ID}  \\ \hline
            \textit{GET\_SEMAPHORE\_ID} & \textit{GET\_PROCESS\_ID} & \textit{GET\_EVENT\_STATUS} & \textit{CREATE\_SAMPLING\_PORT} \\ \hline
            & \textit{UNLOCK\_PREEMPTION}  & \textit{LOCK\_PREEMPTION} &
            \label{tab:apex}
            \end{tabular}
        \end{table}
        
        The method to collect metrics of these calls is the same as the ones illustrated in Section \ref{sec:sfpbench}. Listing \ref{list:apexExample} presents an example of data collection for the CREATE\_BLACKBOARD call. In the example, the collection tags are placed before and after the call, collection the time for two calls (between lines 6-8 and 16-18)  and storing the results.
        
\begin{lstlisting}[firstline=80,lastline=109, language=C, caption=APEX call execution time collection example., label=list:apexExample]
void test_create_blackboard(uint8_t calc_dev) {
  RETURN_CODE_TYPE   err_code;
  DECLARE_TIME_MEASURE()
  INITIALIZE_TIME_VARS("BLACKBOARD");

  INIT_TIME_MEASURE();
  CREATE_BLACKBOARD ("BB", 64, &black_id1, &err_code);
  FINISH_TIME_MEASURE();
  if(err_code == NO_ERROR){VALIDATE_TIME_MEASURE(calc_dev) }
  else{
      PERF_PRINT_STRING("BB Create 1: ");
      PERF_PRINT_NUMBER(err_code);
      PERF_PRINT_EOL();
  }
 
  INIT_TIME_MEASURE();
  CREATE_BLACKBOARD ("BB2", 64, &black_id2, &err_code);
  FINISH_TIME_MEASURE();
  if(err_code == NO_ERROR){ VALIDATE_TIME_MEASURE(calc_dev) }
  else{
      PERF_PRINT_STRING("BB Create 2: ");
      PERF_PRINT_NUMBER(err_code);
      PERF_PRINT_EOL();
  }
}
\end{lstlisting}
    
    \subsection{Complete applications}
    \label{sub:full}
        This group presents full applications, where the collected execution times collect end-to-end metrics: from the beginning of the application execution until its end. Six (6) tests are in this group:
        \begin{itemize}
            \item \textbf{ADPCM} implements the Adaptive Differential Pulse-Code Modulation algorithm \cite{adpcm}, collecting the execution time to perform a given (configurable) number of executions of the entire algorithm for a predefined data-set; 
            \item \textbf{Dijkstra} defines and application that executes the Dijkstra algorithm \cite{dijkstra} for a predefined graph and collects the execution time for a number of executions, configurable during test creation;
            \item \textbf{Sobel} implements the image edge detection Sobel algorithm \cite{sobel} and executes it a given (configurable) number of times, for a predefined input image. The execution time for each execution loop is annotated;
            \item \textbf{APEX APP 1} synthetic application that emulates the execution flow of a real application. It is composed by four (4) processes that use semaphores to synchronize their execution. Each one of the processes is blocked by another one, and when have their access granted, perform a different task (i.e., process two waits on the semaphore, and when it has its access granted, it makes calls to APEX services). The collected time is the end-to-end execution time, before the beginning of the first process and after the end of the fourth process;
            \item \textbf{APEX APP 2} synthetic application that emulates the execution flow of a real application. It is composed by four (4) processes that use events to synchronize their execution. Each one of the processes is blocked by another one, and when have their access granted, perform a different task (i.e., process two waits for an event, and when it detects it, it makes calls to APEX services). The collected time is the end-to-end execution time, before the beginning of the first process and after the end of the fourth process, and;
            \item \textbf{APEX APP 3} synthetic application that emulates the execution flow of a real application. It is composed by two (2) partitions, with two (2) processes each. For each partition, one process is periodic, and the other one is sporadic. The application uses sampling ports to synchronize processes in different partitions, and semaphores to synchronize processes in the same partition. Partition one (1) executes CRC calculations, using randomly created data. The data is them transferred to partition two (2) using a sampling port. Partition two (2) waits on the sampling port for data to be consumed, and using this data, creates matrices and performs the multiplication of these matrices. The collected time is the end-to-end execution time for each partition.
        \end{itemize}
    
\section{Validation platform}
\label{sec:p2020}
    In order to validate the SFPBench framework, it was integrated with an ARINC-653 compliant RTOS and deployed in real hardware. The target RTOS is a popular ARINC-653 compliant RTOS of large deployment in the market\footnote{Due to licensing agreement, the real name of the RTOS will be omitted. From here and on now, in this document, it will be called ARINCRTOS.}, executing on the P2020RDB-PC board \cite{p2020}.
    
    The P2020RDB-PC communications processors deliver high performance per watt for dual- and single-core applications. It uses 45nm process technology fabrication to operate on frequencies up to 1.2 GHz. The main components of the architecture are:
    \begin{itemize}
        \item Dual high-performance Power Architecture e500 cores
        \item 32 KB L1 (32kb for instructions and 32kb for data) and 512KB L2 caches
        \item Three 10/100/1000 Mbps enhanced three-speed Ethernet controllers (eTSECs)
        \item Two PCI Express interfaces
        \item Two Serial RapidIO interfaces
        \item Two SGMII interfaces
        \item Integrated security engine
        \item Dual high-speed USB controllers (USB 2.0)
        \item 1GB DDR3 DRAM
        \item eLBC, eSDHC, Dual I2C, DUART, PIC, DMA, GPIO
    \end{itemize}
    
\begin{figure}[!htb]
      \centering
      \includegraphics[width=12cm]{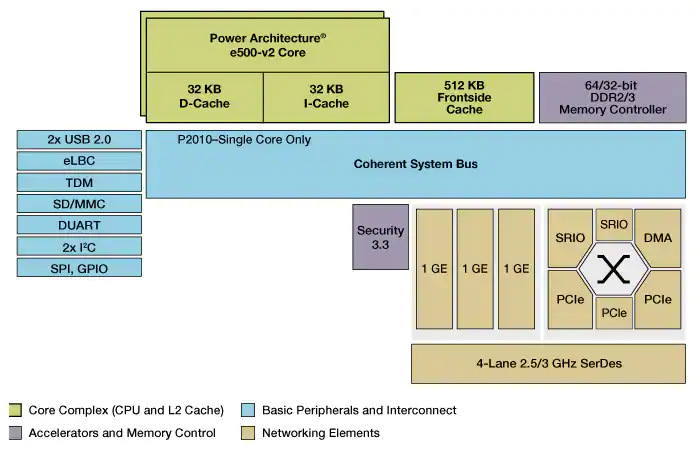}
      \caption{P2020RDB-PC Block Diagram\cite{p2020_img}.}
      \label{fig:p2020}
    \end{figure}

    Figure \ref{fig:p2020} illustrates the P2020RDB-PC block diagram of the board architecture.
    
    ARINCRTOS is configured to execute using top frequency of the processor (1.2 GHz), executing at full speed. L1 cache is enabled for both data and instructions and cache L2 is configure as SRAM, to improve the performance of the RTOS kernel. The compiler (GCC) is configure to enable optimizations, using -O2 optimization flag.
    
\section{SFPBench results exploration}
\label{sec:results}
    To illustrate the kind of results that are generated by the usage of SFPBench framework, the entire set of applications was executed with ARINCRTOS, on the P2020RDB-PC board. The next tables present all the data generated by the framework, divided by application and for each application, the execution times in clock ticks and microseconds are illustrated. Further, best-case execution time (BCET), worst-case execution time (WCET) and average execution time are given. Table \ref{tab:grey_results} presents the obtained results for the grey-box applications, followed by Table \ref{tab:apex_results}, which presents the APEX calls obtained results. Finally, Table \ref{tab:app_results} presents the results obtained for the complete applications. 
 
        \begin{table}[!htb!]
        \centering
        \caption{Grey-box applications execution times.}
        \small
        \begin{tabular}{l|c|c|c|c|c|c|c|}
        \cline{2-8}
        \multicolumn{1}{l|}{}                                                  & \multicolumn{3}{c|}{\textbf{Time (ticks)}}                                                                                                             & \multicolumn{4}{c|}{\textbf{Time (us)}}                                                                                                                               \\ \cline{2-8} 
        \multicolumn{1}{l|}{}                                                  & \textbf{BCET}                                    & \textbf{WCET}                                    & \textbf{Average}                                 & \textbf{BCET}                                  & \textbf{WCET}                         & \textbf{Average}                      & \textbf{STD Dev.}                    \\ \hline
        \multicolumn{1}{|c|}{{\color[HTML]{215967} \textbf{Process Switch}}}   & {\color[HTML]{215967} \textit{\textbf{113.00}}}  & {\color[HTML]{215967} \textit{\textbf{1290.00}}} & {\color[HTML]{215967} \textit{\textbf{113.00}}}  & {\color[HTML]{215967} \textit{\textbf{1.50}}}  & {\color[HTML]{215967} \textbf{17.20}} & {\color[HTML]{215967} \textbf{1.63}}  & {\color[HTML]{215967} \textbf{0.85}} \\ \hline
        \multicolumn{1}{|c|}{{\color[HTML]{215967} \textbf{Mutex Acquire}}}      & {\color[HTML]{215967} \textit{\textbf{75.00}}}   & {\color[HTML]{215967} \textit{\textbf{947.00}}}  & {\color[HTML]{215967} \textit{\textbf{76.00}}}   & {\color[HTML]{215967} \textit{\textbf{1.00}}}  & {\color[HTML]{215967} \textbf{12.62}} & {\color[HTML]{215967} \textbf{1.05}}  & {\color[HTML]{215967} \textbf{1.23}} \\ \hline
        \multicolumn{1}{|c|}{{\color[HTML]{215967} \textbf{Mutex Release}}}    & {\color[HTML]{215967} \textit{\textbf{79.00}}}   & {\color[HTML]{215967} \textit{\textbf{1128.00}}} & {\color[HTML]{215967} \textit{\textbf{79.00}}}   & {\color[HTML]{215967} \textit{\textbf{1.05}}}  & {\color[HTML]{215967} \textbf{15.04}} & {\color[HTML]{215967} \textbf{1.05}}  & {\color[HTML]{215967} \textbf{0.17}} \\ \hline
        \multicolumn{1}{|c|}{{\color[HTML]{215967} \textbf{Mutex Acquire 2}}}    & {\color[HTML]{215967} \textit{\textbf{27.00}}}   & {\color[HTML]{215967} \textit{\textbf{112.00}}}  & {\color[HTML]{215967} \textit{\textbf{27.00}}}   & {\color[HTML]{215967} \textit{\textbf{0.36}}}  & {\color[HTML]{215967} \textbf{1.49}}  & {\color[HTML]{215967} \textbf{0.36}}  & {\color[HTML]{215967} \textbf{0.00}} \\ \hline
        \multicolumn{1}{|c|}{{\color[HTML]{215967} \textbf{Mutex Release 2}}}    & {\color[HTML]{215967} \textit{\textbf{29.00}}}   & {\color[HTML]{215967} \textit{\textbf{203.00}}}  & {\color[HTML]{215967} \textit{\textbf{29.00}}}   & {\color[HTML]{215967} \textit{\textbf{0.38}}}  & {\color[HTML]{215967} \textbf{2.70}}  & {\color[HTML]{215967} \textbf{0.38}}  & {\color[HTML]{215967} \textbf{0.01}} \\ \hline
        \multicolumn{1}{|c|}{{\color[HTML]{215967} \textbf{Mutex Workload}}}   & {\color[HTML]{215967} \textit{\textbf{142.00}}}  & {\color[HTML]{215967} \textit{\textbf{181.00}}}  & {\color[HTML]{215967} \textit{\textbf{146.00}}}  & {\color[HTML]{215967} \textit{\textbf{1.89}}}  & {\color[HTML]{215967} \textbf{2.41}}  & {\color[HTML]{215967} \textbf{1.94}}  & {\color[HTML]{215967} \textbf{0.01}} \\ \hline
        \multicolumn{1}{|c|}{{\color[HTML]{215967} \textbf{Sem Wait}}}         & {\color[HTML]{215967} \textit{\textbf{118.00}}}  & {\color[HTML]{215967} \textit{\textbf{831.00}}}  & {\color[HTML]{215967} \textit{\textbf{119.00}}}  & {\color[HTML]{215967} \textit{\textbf{1.57}}}  & {\color[HTML]{215967} \textbf{11.08}} & {\color[HTML]{215967} \textbf{1.58}}  & {\color[HTML]{215967} \textbf{0.17}} \\ \hline
        \multicolumn{1}{|c|}{{\color[HTML]{215967} \textbf{Sem Signal}}}       & {\color[HTML]{215967} \textit{\textbf{75.00}}}   & {\color[HTML]{215967} \textit{\textbf{879.00}}}  & {\color[HTML]{215967} \textit{\textbf{75.00}}}   & {\color[HTML]{215967} \textit{\textbf{1.00}}}  & {\color[HTML]{215967} \textbf{11.72}} & {\color[HTML]{215967} \textbf{1.00}}  & {\color[HTML]{215967} \textbf{0.01}} \\ \hline
        \multicolumn{1}{|c|}{{\color[HTML]{215967} \textbf{Priority Sem}}}     & {\color[HTML]{215967} \textit{\textbf{44.00}}}   & {\color[HTML]{215967} \textit{\textbf{1213.00}}} & {\color[HTML]{215967} \textit{\textbf{52.00}}}   & {\color[HTML]{215967} \textit{\textbf{0.58}}}  & {\color[HTML]{215967} \textbf{16.17}} & {\color[HTML]{215967} \textbf{0.69}}  & {\color[HTML]{215967} \textbf{0.01}} \\ \hline
        \multicolumn{1}{|c|}{{\color[HTML]{215967} \textbf{Sem Signal 2}}}     & {\color[HTML]{215967} \textit{\textbf{28.00}}}   & {\color[HTML]{215967} \textit{\textbf{128.00}}}  & {\color[HTML]{215967} \textit{\textbf{39.00}}}   & {\color[HTML]{215967} \textit{\textbf{0.37}}}  & {\color[HTML]{215967} \textbf{1.70}}  & {\color[HTML]{215967} \textbf{0.52}}  & {\color[HTML]{215967} \textbf{0.29}} \\ \hline
        \multicolumn{1}{|c|}{{\color[HTML]{215967} \textbf{Sem Wait 2}}}       & {\color[HTML]{215967} \textit{\textbf{34.00}}}   & {\color[HTML]{215967} \textit{\textbf{196.00}}}  & {\color[HTML]{215967} \textit{\textbf{94.00}}}   & {\color[HTML]{215967} \textit{\textbf{0.45}}}  & {\color[HTML]{215967} \textbf{2.60}}  & {\color[HTML]{215967} \textbf{1.25}}  & {\color[HTML]{215967} \textbf{0.81}} \\ \hline
        \multicolumn{1}{|c|}{{\color[HTML]{215967} \textbf{Sem Workload}}}     & {\color[HTML]{215967} \textit{\textbf{138.00}}}  & {\color[HTML]{215967} \textit{\textbf{301.00}}}  & {\color[HTML]{215967} \textit{\textbf{456.00}}}  & {\color[HTML]{215967} \textit{\textbf{1.84}}}  & {\color[HTML]{215967} \textbf{6.08}}  & {\color[HTML]{215967} \textbf{4.01}}  & {\color[HTML]{215967} \textbf{2.11}} \\ \hline
        \multicolumn{1}{|c|}{{\color[HTML]{215967} \textbf{Partition Switch}}} & {\color[HTML]{215967} \textit{\textbf{1662.00}}} & {\color[HTML]{215967} \textit{\textbf{3056.00}}} & {\color[HTML]{215967} \textit{\textbf{1682.00}}} & {\color[HTML]{215967} \textit{\textbf{22.16}}} & {\color[HTML]{215967} \textbf{40.74}} & {\color[HTML]{215967} \textbf{22.44}} & {\color[HTML]{215967} \textbf{5.68}} \\ \hline
        \label{tab:grey_results}
        \end{tabular}
        \end{table}

\begin{table}[!htb!]
    \centering
    \caption{APEX calls execution times.}
    \small
\begin{tabular}{l|c|c|c|c|c|c|c|}
\cline{2-8}
\multicolumn{1}{l|}{{\color[HTML]{215967} \textbf{}}}                    & \multicolumn{3}{c|}{\textbf{Time (ticks)}}                                                                                                             & \multicolumn{4}{c|}{\textbf{Time (us)}}                                                                                                                                       \\ \cline{2-8} 
\multicolumn{1}{l|}{{\color[HTML]{215967} \textbf{}}}                    & {\color[HTML]{215967} \textbf{BCET}}             & {\color[HTML]{215967} \textbf{WCET}}             & {\color[HTML]{215967} \textbf{Average}}          & {\color[HTML]{215967} \textbf{BCET}}            & {\color[HTML]{215967} \textbf{WCET}}   & {\color[HTML]{215967} \textbf{Average}} & {\color[HTML]{215967} \textbf{STD Dev.}} \\ \hline
\multicolumn{1}{|c|}{{\color[HTML]{215967} \textbf{INIT\_PROCESS1}}}     & {\color[HTML]{215967} \textit{\textbf{8054.00}}} & {\color[HTML]{215967} \textit{\textbf{8054.00}}} & {\color[HTML]{215967} \textit{\textbf{8054.00}}} & {\color[HTML]{215967} \textit{\textbf{107.38}}} & {\color[HTML]{215967} \textbf{107.38}} & {\color[HTML]{215967} \textbf{107.38}}  & {\color[HTML]{215967} \textbf{0.00}}     \\ \hline
\multicolumn{1}{|c|}{{\color[HTML]{215967} \textbf{INIT\_PROCESS2}}}     & {\color[HTML]{215967} \textit{\textbf{2796.00}}} & {\color[HTML]{215967} \textit{\textbf{2796.00}}} & {\color[HTML]{215967} \textit{\textbf{2796.00}}} & {\color[HTML]{215967} \textit{\textbf{37.27}}}  & {\color[HTML]{215967} \textbf{37.27}}  & {\color[HTML]{215967} \textbf{37.27}}   & {\color[HTML]{215967} \textbf{0.00}}     \\ \hline
\multicolumn{1}{|c|}{{\color[HTML]{215967} \textbf{SEMAPHORE\_CREATE}}}  & {\color[HTML]{215967} \textit{\textbf{342.00}}}  & {\color[HTML]{215967} \textit{\textbf{2555.00}}} & {\color[HTML]{215967} \textit{\textbf{1004.00}}} & {\color[HTML]{215967} \textit{\textbf{4.56}}}   & {\color[HTML]{215967} \textbf{34.06}}  & {\color[HTML]{215967} \textbf{13.38}}   & {\color[HTML]{215967} \textbf{0.00}}     \\ \hline
\multicolumn{1}{|c|}{{\color[HTML]{215967} \textbf{BUFFER\_CREATE}}}     & {\color[HTML]{215967} \textit{\textbf{335.00}}}  & {\color[HTML]{215967} \textit{\textbf{490.00}}}  & {\color[HTML]{215967} \textit{\textbf{388.00}}}  & {\color[HTML]{215967} \textit{\textbf{4.46}}}   & {\color[HTML]{215967} \textbf{6.53}}   & {\color[HTML]{215967} \textbf{5.18}}    & {\color[HTML]{215967} \textbf{0.00}}     \\ \hline
\multicolumn{1}{|c|}{{\color[HTML]{215967} \textbf{BB\_CREATE}}}         & {\color[HTML]{215967} \textit{\textbf{284.00}}}  & {\color[HTML]{215967} \textit{\textbf{415.00}}}  & {\color[HTML]{215967} \textit{\textbf{349.00}}}  & {\color[HTML]{215967} \textit{\textbf{3.78}}}   & {\color[HTML]{215967} \textbf{5.53}}   & {\color[HTML]{215967} \textbf{4.65}}    & {\color[HTML]{215967} \textbf{0.00}}     \\ \hline
\multicolumn{1}{|c|}{{\color[HTML]{215967} \textbf{CREATE\_EVENT}}}      & {\color[HTML]{215967} \textit{\textbf{329.00}}}  & {\color[HTML]{215967} \textit{\textbf{594.00}}}  & {\color[HTML]{215967} \textit{\textbf{407.00}}}  & {\color[HTML]{215967} \textit{\textbf{4.38}}}   & {\color[HTML]{215967} \textbf{7.92}}   & {\color[HTML]{215967} \textbf{5.43}}    & {\color[HTML]{215967} \textbf{0.00}}     \\ \hline
\multicolumn{1}{|c|}{{\color[HTML]{215967} \textbf{PARTITION\_STATUS}}}  & {\color[HTML]{215967} \textit{\textbf{22.00}}}   & {\color[HTML]{215967} \textit{\textbf{169.00}}}  & {\color[HTML]{215967} \textit{\textbf{58.00}}}   & {\color[HTML]{215967} \textit{\textbf{0.29}}}   & {\color[HTML]{215967} \textbf{2.25}}   & {\color[HTML]{215967} \textbf{0.78}}    & {\color[HTML]{215967} \textbf{0.84}}     \\ \hline
\multicolumn{1}{|c|}{{\color[HTML]{215967} \textbf{PREEMPTION\_LOCK}}}   & {\color[HTML]{215967} \textit{\textbf{111.00}}}  & {\color[HTML]{215967} \textit{\textbf{298.00}}}  & {\color[HTML]{215967} \textit{\textbf{157.00}}}  & {\color[HTML]{215967} \textit{\textbf{1.48}}}   & {\color[HTML]{215967} \textbf{3.97}}   & {\color[HTML]{215967} \textbf{2.10}}    & {\color[HTML]{215967} \textbf{1.07}}     \\ \hline
\multicolumn{1}{|c|}{{\color[HTML]{215967} \textbf{PREEMPTION\_UNLOCK}}} & {\color[HTML]{215967} \textbf{129.00}}           & {\color[HTML]{215967} \textbf{130.00}}           & {\color[HTML]{215967} \textbf{129.00}}           & {\color[HTML]{215967} \textbf{1.72}}            & {\color[HTML]{215967} \textbf{1.73}}   & {\color[HTML]{215967} \textbf{1.72}}    & {\color[HTML]{215967} \textbf{0.05}}     \\ \hline
\multicolumn{1}{|c|}{{\color[HTML]{215967} \textbf{DISPLAY\_BB(16)}}}    & {\color[HTML]{215967} \textbf{31.00}}            & {\color[HTML]{215967} \textbf{114.00}}           & {\color[HTML]{215967} \textbf{51.00}}            & {\color[HTML]{215967} \textbf{0.41}}            & {\color[HTML]{215967} \textbf{1.52}}   & {\color[HTML]{215967} \textbf{0.68}}    & {\color[HTML]{215967} \textbf{0.47}}     \\ \hline
\multicolumn{1}{|c|}{{\color[HTML]{215967} \textbf{READ\_BB(16)}}}       & {\color[HTML]{215967} \textbf{59.00}}            & {\color[HTML]{215967} \textbf{147.00}}           & {\color[HTML]{215967} \textbf{83.00}}            & {\color[HTML]{215967} \textbf{0.78}}            & {\color[HTML]{215967} \textbf{1.96}}   & {\color[HTML]{215967} \textbf{1.11}}    & {\color[HTML]{215967} \textbf{0.48}}     \\ \hline
\multicolumn{1}{|c|}{{\color[HTML]{215967} \textbf{DISPLAY\_BB(64)}}}    & {\color[HTML]{215967} \textbf{34.00}}            & {\color[HTML]{215967} \textbf{43.00}}            & {\color[HTML]{215967} \textbf{35.00}}            & {\color[HTML]{215967} \textbf{0.45}}            & {\color[HTML]{215967} \textbf{0.57}}   & {\color[HTML]{215967} \textbf{0.48}}    & {\color[HTML]{215967} \textbf{0.05}}     \\ \hline
\multicolumn{1}{|c|}{{\color[HTML]{215967} \textbf{READ\_BB(64)}}}       & {\color[HTML]{215967} \textbf{63.00}}            & {\color[HTML]{215967} \textbf{65.00}}            & {\color[HTML]{215967} \textbf{63.00}}            & {\color[HTML]{215967} \textbf{0.84}}            & {\color[HTML]{215967} \textbf{0.86}}   & {\color[HTML]{215967} \textbf{0.85}}    & {\color[HTML]{215967} \textbf{0.01}}     \\ \hline
\multicolumn{1}{|c|}{{\color[HTML]{215967} \textbf{SEND\_BUFFER(16)}}}   & {\color[HTML]{215967} \textbf{90.00}}            & {\color[HTML]{215967} \textbf{226.00}}           & {\color[HTML]{215967} \textbf{134.00}}           & {\color[HTML]{215967} \textbf{1.20}}            & {\color[HTML]{215967} \textbf{3.01}}   & {\color[HTML]{215967} \textbf{1.79}}    & {\color[HTML]{215967} \textbf{0.73}}     \\ \hline
\multicolumn{1}{|c|}{{\color[HTML]{215967} \textbf{SEMAPHORE\_SIGNAL}}}  & {\color[HTML]{215967} \textbf{30.00}}            & {\color[HTML]{215967} \textbf{79.00}}            & {\color[HTML]{215967} \textbf{42.00}}            & {\color[HTML]{215967} \textbf{0.40}}            & {\color[HTML]{215967} \textbf{1.05}}   & {\color[HTML]{215967} \textbf{0.56}}    & {\color[HTML]{215967} \textbf{0.27}}     \\ \hline
\multicolumn{1}{|c|}{{\color[HTML]{215967} \textbf{PROCESS\_PRIORITY}}}  & {\color[HTML]{215967} \textbf{65.00}}            & {\color[HTML]{215967} \textbf{209.00}}           & {\color[HTML]{215967} \textbf{101.00}}           & {\color[HTML]{215967} \textbf{0.86}}            & {\color[HTML]{215967} \textbf{2.78}}   & {\color[HTML]{215967} \textbf{1.34}}    & {\color[HTML]{215967} \textbf{0.82}}     \\ \hline
\multicolumn{1}{|c|}{{\color[HTML]{215967} \textbf{MY\_ID}}}             & {\color[HTML]{215967} \textbf{43.00}}            & {\color[HTML]{215967} \textbf{80.00}}            & {\color[HTML]{215967} \textbf{52.00}}            & {\color[HTML]{215967} \textbf{0.57}}            & {\color[HTML]{215967} \textbf{1.06}}   & {\color[HTML]{215967} \textbf{0.69}}    & {\color[HTML]{215967} \textbf{0.21}}     \\ \hline
\multicolumn{1}{|c|}{{\color[HTML]{215967} \textbf{PROCESS\_ID}}}        & {\color[HTML]{215967} \textbf{34.00}}            & {\color[HTML]{215967} \textbf{40.00}}            & {\color[HTML]{215967} \textbf{36.00}}            & {\color[HTML]{215967} \textbf{0.45}}            & {\color[HTML]{215967} \textbf{0.53}}   & {\color[HTML]{215967} \textbf{0.48}}    & {\color[HTML]{215967} \textbf{0.03}}     \\ \hline
\multicolumn{1}{|c|}{{\color[HTML]{215967} \textbf{PROCESS\_STATUS}}}    & {\color[HTML]{215967} \textbf{67.00}}            & {\color[HTML]{215967} \textbf{166.00}}           & {\color[HTML]{215967} \textbf{91.00}}            & {\color[HTML]{215967} \textbf{0.89}}            & {\color[HTML]{215967} \textbf{2.21}}   & {\color[HTML]{215967} \textbf{1.22}}    & {\color[HTML]{215967} \textbf{0.56}}     \\ \hline
\multicolumn{1}{|c|}{{\color[HTML]{215967} \textbf{SEMAPHORE\_ID}}}      & {\color[HTML]{215967} \textbf{32.00}}            & {\color[HTML]{215967} \textbf{59.00}}            & {\color[HTML]{215967} \textbf{44.00}}            & {\color[HTML]{215967} \textbf{0.42}}            & {\color[HTML]{215967} \textbf{0.78}}   & {\color[HTML]{215967} \textbf{0.59}}    & {\color[HTML]{215967} \textbf{0.13}}     \\ \hline
\multicolumn{1}{|c|}{{\color[HTML]{215967} \textbf{SEMAPHORE\_STATUS}}}  & {\color[HTML]{215967} \textbf{38.00}}            & {\color[HTML]{215967} \textbf{112.00}}           & {\color[HTML]{215967} \textbf{57.00}}            & {\color[HTML]{215967} \textbf{0.50}}            & {\color[HTML]{215967} \textbf{1.49}}   & {\color[HTML]{215967} \textbf{0.76}}    & {\color[HTML]{215967} \textbf{0.42}}     \\ \hline
\multicolumn{1}{|c|}{{\color[HTML]{215967} \textbf{EVENT\_SET}}}         & {\color[HTML]{215967} \textbf{29.00}}            & {\color[HTML]{215967} \textbf{75.00}}            & {\color[HTML]{215967} \textbf{50.00}}            & {\color[HTML]{215967} \textbf{0.38}}            & {\color[HTML]{215967} \textbf{1.00}}   & {\color[HTML]{215967} \textbf{0.68}}    & {\color[HTML]{215967} \textbf{0.29}}     \\ \hline
\multicolumn{1}{|c|}{{\color[HTML]{215967} \textbf{EVENT\_ID}}}          & {\color[HTML]{215967} \textbf{42.00}}            & {\color[HTML]{215967} \textbf{105.00}}           & {\color[HTML]{215967} \textbf{65.00}}            & {\color[HTML]{215967} \textbf{0.56}}            & {\color[HTML]{215967} \textbf{1.40}}   & {\color[HTML]{215967} \textbf{0.87}}    & {\color[HTML]{215967} \textbf{0.31}}     \\ \hline
\multicolumn{1}{|c|}{{\color[HTML]{215967} \textbf{EVENT\_STATUS}}}      & {\color[HTML]{215967} \textbf{30.00}}            & {\color[HTML]{215967} \textbf{104.00}}           & {\color[HTML]{215967} \textbf{48.00}}            & {\color[HTML]{215967} \textbf{0.40}}            & {\color[HTML]{215967} \textbf{1.38}}   & {\color[HTML]{215967} \textbf{0.64}}    & {\color[HTML]{215967} \textbf{0.42}}     \\ \hline
\multicolumn{1}{|c|}{{\color[HTML]{215967} \textbf{CREATE\_SAMPLING}}}   & {\color[HTML]{215967} \textbf{752.00}}           & {\color[HTML]{215967} \textbf{2671.00}}          & {\color[HTML]{215967} \textbf{1436.00}}          & {\color[HTML]{215967} \textbf{10.02}}           & {\color[HTML]{215967} \textbf{35.61}}  & {\color[HTML]{215967} \textbf{19.14}}   & {\color[HTML]{215967} \textbf{0.00}}     \\ \hline
\multicolumn{1}{|c|}{{\color[HTML]{215967} \textbf{CREATE\_QUEUE}}}      & {\color[HTML]{215967} \textbf{1262.00}}          & {\color[HTML]{215967} \textbf{2439.00}}          & {\color[HTML]{215967} \textbf{1683.00}}          & {\color[HTML]{215967} \textbf{16.82}}           & {\color[HTML]{215967} \textbf{32.52}}  & {\color[HTML]{215967} \textbf{22.45}}   & {\color[HTML]{215967} \textbf{0.00}}     \\ \hline
\multicolumn{1}{|c|}{{\color[HTML]{215967} \textbf{SAMPLING\_ID}}}       & {\color[HTML]{215967} \textbf{34.00}}            & {\color[HTML]{215967} \textbf{873.00}}           & {\color[HTML]{215967} \textbf{279.00}}           & {\color[HTML]{215967} \textbf{0.45}}            & {\color[HTML]{215967} \textbf{11.64}}  & {\color[HTML]{215967} \textbf{3.73}}    & {\color[HTML]{215967} \textbf{4.60}}     \\ \hline
\multicolumn{1}{|c|}{{\color[HTML]{215967} \textbf{SAMPLING\_STATUS}}}   & {\color[HTML]{215967} \textbf{17.00}}            & {\color[HTML]{215967} \textbf{45.00}}            & {\color[HTML]{215967} \textbf{26.00}}            & {\color[HTML]{215967} \textbf{0.22}}            & {\color[HTML]{215967} \textbf{0.60}}   & {\color[HTML]{215967} \textbf{0.34}}    & {\color[HTML]{215967} \textbf{0.14}}     \\ \hline
\multicolumn{1}{|c|}{{\color[HTML]{215967} \textbf{QUEUE\_STATUS}}}      & {\color[HTML]{215967} \textbf{184.00}}           & {\color[HTML]{215967} \textbf{575.00}}           & {\color[HTML]{215967} \textbf{295.00}}           & {\color[HTML]{215967} \textbf{2.45}}            & {\color[HTML]{215967} \textbf{7.66}}   & {\color[HTML]{215967} \textbf{3.94}}    & {\color[HTML]{215967} \textbf{2.15}}     \\ \hline
\multicolumn{1}{|c|}{{\color[HTML]{215967} \textbf{QUEUE\_ID}}}          & {\color[HTML]{215967} \textbf{28.00}}            & {\color[HTML]{215967} \textbf{35.00}}            & {\color[HTML]{215967} \textbf{30.00}}            & {\color[HTML]{215967} \textbf{0.37}}            & {\color[HTML]{215967} \textbf{0.46}}   & {\color[HTML]{215967} \textbf{0.41}}    & {\color[HTML]{215967} \textbf{0.03}}     \\ \hline
\multicolumn{1}{|c|}{{\color[HTML]{215967} \textbf{QUEUE\_WRITE}}}       & {\color[HTML]{215967} \textbf{841.00}}           & {\color[HTML]{215967} \textbf{1681.00}}          & {\color[HTML]{215967} \textbf{1261.00}}          & {\color[HTML]{215967} \textbf{11.21}}           & {\color[HTML]{215967} \textbf{22.41}}  & {\color[HTML]{215967} \textbf{16.81}}   & {\color[HTML]{215967} \textbf{5.60}}     \\ \hline
\multicolumn{1}{|c|}{{\color[HTML]{215967} \textbf{SAMPLING\_WRITE}}}    & {\color[HTML]{215967} \textbf{335.00}}           & {\color[HTML]{215967} \textbf{757.00}}           & {\color[HTML]{215967} \textbf{546.00}}           & {\color[HTML]{215967} \textbf{4.46}}            & {\color[HTML]{215967} \textbf{10.09}}  & {\color[HTML]{215967} \textbf{7.27}}    & {\color[HTML]{215967} \textbf{2.81}}     \\ \hline
\multicolumn{1}{|c|}{{\color[HTML]{215967} \textbf{SAMPLING\_READ}}}     & {\color[HTML]{215967} \textbf{314.00}}           & {\color[HTML]{215967} \textbf{636.00}}           & {\color[HTML]{215967} \textbf{475.00}}           & {\color[HTML]{215967} \textbf{4.18}}            & {\color[HTML]{215967} \textbf{8.48}}   & {\color[HTML]{215967} \textbf{6.33}}    & {\color[HTML]{215967} \textbf{2.14}}     \\ \hline
\label{tab:apex_results}
\end{tabular}
\end{table}
        
\begin{table}[!htb!]
       \centering
        \caption{Complete applications execution times.}
        \small
\begin{tabular}{l|c|c|c|c|c|c|c|}
\cline{2-8}
{\color[HTML]{215967} \textbf{}}                                         & \multicolumn{3}{c|}{\textbf{Time (ticks)}}                                                                                                                         & \multicolumn{4}{c|}{\textbf{Time (us)}}                                                                                                                                                  \\ \cline{2-8} 
\multicolumn{1}{l|}{{\color[HTML]{215967} \textbf{}}}                    & {\color[HTML]{215967} \textbf{BCET}}                 & {\color[HTML]{215967} \textbf{WCET}}                 & {\color[HTML]{215967} \textbf{Average}}              & {\color[HTML]{215967} \textbf{BCET}}                & {\color[HTML]{215967} \textbf{WCET}}       & {\color[HTML]{215967} \textbf{Average}}    & {\color[HTML]{215967} \textbf{STD Dev.}} \\ \hline
\multicolumn{1}{|c|}{{\color[HTML]{215967} \textbf{SOBEL}}}              & {\color[HTML]{215967} \textit{\textbf{393084.00}}}   & {\color[HTML]{215967} \textit{\textbf{393834.00}}}   & {\color[HTML]{215967} \textit{\textbf{393204.00}}}   & {\color[HTML]{215967} \textit{\textbf{5240.64}}}    & {\color[HTML]{215967} \textbf{5251.12}}    & {\color[HTML]{215967} \textbf{5242.86}}    & {\color[HTML]{215967} \textbf{2.89}}     \\ \hline
\multicolumn{1}{|c|}{{\color[HTML]{215967} \textbf{ADPCM}}}              & {\color[HTML]{215967} \textit{\textbf{26724.00}}}    & {\color[HTML]{215967} \textit{\textbf{29304.00}}}    & {\color[HTML]{215967} \textit{\textbf{27360.00}}}    & {\color[HTML]{215967} \textit{\textbf{356.32}}}     & {\color[HTML]{215967} \textbf{390.72}}     & {\color[HTML]{215967} \textbf{364.85}}     & {\color[HTML]{215967} \textbf{8.53}}     \\ \hline
\multicolumn{1}{|c|}{{\color[HTML]{215967} \textbf{DIJKSTRA}}}           & {\color[HTML]{215967} \textit{\textbf{2322.00}}}     & {\color[HTML]{215967} \textit{\textbf{4364.00}}}     & {\color[HTML]{215967} \textit{\textbf{2488.00}}}     & {\color[HTML]{215967} \textit{\textbf{30.95}}}      & {\color[HTML]{215967} \textbf{58.18}}      & {\color[HTML]{215967} \textbf{33.38}}      & {\color[HTML]{215967} \textbf{5.78}}     \\ \hline
\multicolumn{1}{|c|}{{\color[HTML]{215967} \textbf{APEX APP 1}}}         & {\color[HTML]{215967} \textit{\textbf{578.00}}}      & {\color[HTML]{215967} \textit{\textbf{2845.00}}}     & {\color[HTML]{215967} \textit{\textbf{629.00}}}      & {\color[HTML]{215967} \textit{\textbf{7.70}}}       & {\color[HTML]{215967} \textbf{37.93}}      & {\color[HTML]{215967} \textbf{8.65}}       & {\color[HTML]{215967} \textbf{4.12}}     \\ \hline
\multicolumn{1}{|c|}{{\color[HTML]{215967} \textbf{APEX APP 2}}}         & {\color[HTML]{215967} \textit{\textbf{838.00}}}      & {\color[HTML]{215967} \textit{\textbf{2698.00}}}     & {\color[HTML]{215967} \textit{\textbf{919.00}}}      & {\color[HTML]{215967} \textit{\textbf{11.17}}}      & {\color[HTML]{215967} \textbf{35.97}}      & {\color[HTML]{215967} \textbf{12.53}}      & {\color[HTML]{215967} \textbf{4.29}}     \\ \hline
\multicolumn{1}{|c|}{{\color[HTML]{215967} \textbf{SAMPLE APEX APP  A}}} & {\color[HTML]{215967} \textit{\textbf{54469113.00}}} & {\color[HTML]{215967} \textit{\textbf{54469113.00}}} & {\color[HTML]{215967} \textit{\textbf{54469113.00}}} & {\color[HTML]{215967} \textit{\textbf{726197.73}}}  & {\color[HTML]{215967} \textbf{726197.73}}  & {\color[HTML]{215967} \textbf{726197.73}}  & {\color[HTML]{215967} \textbf{0.00}}     \\ \hline
\multicolumn{1}{|c|}{{\color[HTML]{215967} \textbf{SAMPLE APEX APP  B}}} & {\color[HTML]{215967} \textit{\textbf{36710197.00}}} & {\color[HTML]{215967} \textit{\textbf{36710197.00}}} & {\color[HTML]{215967} \textit{\textbf{36710197.00}}} & {\color[HTML]{215967} \textit{\textbf{489469.29}}}  & {\color[HTML]{215967} \textbf{489469.29}}  & {\color[HTML]{215967} \textbf{489469.29}}  & {\color[HTML]{215967} \textbf{0.00}}     \\ \hline
\multicolumn{1}{|c|}{{\color[HTML]{215967} \textbf{SAMPLE APEX TOTAL}}}  & {\color[HTML]{215967} \textit{\textbf{91179310.00}}} & {\color[HTML]{215967} \textit{\textbf{91179310.00}}} & {\color[HTML]{215967} \textit{\textbf{91179310.00}}} & {\color[HTML]{215967} \textit{\textbf{1215667.03}}} & {\color[HTML]{215967} \textbf{1215667.03}} & {\color[HTML]{215967} \textbf{1215667.03}} & {\color[HTML]{215967} \textbf{0.00}}     \\ \hline
\label{tab:app_results}
\end{tabular}
\end{table}    
        
        \newpage Tables \ref{tab:grey_results}, \ref{tab:apex_results} and \ref{tab:app_results} show that SFPBench enables designers to collect a rich gamma of information regarding execution metrics of the system. Its usage is simple, from a deployment perspective and the results publishing are automated. The provided tests assess the main aspects of an ARINC-653 compliant RTOS, enabling avionic application engineers to optimize their designs, relying on the collected results.
        
        To verify the accuracy of the results collected with SFPBench, the same application execution metrics were assessed using a tracing probe, with timing capabilities. When compared to the traced data, the error rate associated with the higher level of abstraction of SFPBench is up to 2\%.
        
\section{Conclusion and Future work}
\label{sec:conclusion}
    ARINC-653 RTOSs are a particular class of RTOSs, where time and space of different applications should be isolated. On the opposite of regular RTOS that have available different benchmark frameworks, ARINC-653 compliant RTOSs do not have any. This document presented an open-source \cite{github_sff}, collaborative benchmark framework targeting ARINC-653 compliant RTOSs, the SFPBench. The goal of SFPBench is to establish common performance metrics, and from a set of tests define reference points for comparison between systems. The suite is easily portable to different RTOSs, relying on a simple porting layer to enable its usage. 
    
    Possible future work comprise the deployment of additional test applications, stressing different aspects of the systems, such as boot-up time and memory management. 
    
\bibliographystyle{unsrt}  
\bibliography{SFPBench}

\begin{thebibliography}{10}

\bibitem{arinc653}
ARINC-653.
\newblock {ARINC 653P0-1 Avionics Application Software Standard Interface, Part
  0, ARINC 653}, Aug 2015.

\bibitem{wcet}
J.~Abella, D.~Hardy, I.~Puaut, E.~Quinones, and F.J. Cazorla.
\newblock {On the Comparison of Deterministic and Probabilistic WCET Estimation
  Techniques}.
\newblock pages 266 -- 75, Piscataway, NJ, USA, 2014//.

\bibitem{vxworks}
Paul Parkinson.
\newblock {Update on using multicore processors with a commercial ARINC 653
  implementation}.
\newblock 04 2017.

\bibitem{pikeos}
{Kaiser, Robert and Wagner, Stephan}.
\newblock {Evolution of the PikeOS Microkernel}.
\newblock 02 2007.

\bibitem{integrity}
M.~A. Griglock\textit{et al}.
\newblock Time-variant scheduling of affinity groups on a multi-core processor,
  Nov 2012.

\bibitem{github_sff}
{SFPBench: ARINC-653 Benchmark Suite}.
\newblock {https://github.com/HESL-polymtl/benchmark}, 05 2020.

\bibitem{4391842}
C.~B. Watkins and R.~Walter.
\newblock Transitioning from federated avionics architectures to integrated
  modular avionics.
\newblock In {\em 2007 IEEE/AIAA 26th Digital Avionics Systems Conference},
  pages 2.A.1--1--2.A.1--10, Oct 2007.

\bibitem{bench1}
{J. Peleska \textit{et al}}.
\newblock {A Real-world Benchmark Model for Testing Concurrent Real-time
  Systems in the Automotive Domain}.
\newblock In {\em Proceedings of the 23rd IFIP WG 6.1 International Conference
  on Testing Software and Systems}, ICTSS'11. Springer-Verlag, 2011.

\bibitem{bench2}
{C.D. Spradling}.
\newblock {SPEC CPU2006 Benchmark Tools}.
\newblock {\em SIGARCH Computer Architecture News}, 35, March 2007.

\bibitem{bench3}
{SNU Real-Time Benchmarks}.
\newblock {http://www.cprover.org/goto-cc/examples/snu.html}, 12 2019.

\bibitem{bench4}
{THREAD Metric Benchmark Test Suite}.
\newblock {shorturl.at/fgmJU},
  12 2019.

\bibitem{adpcm}
K.C. Pohlmann.
\newblock {\em {Principles of Digital Audio, Sixth Edition}}.
\newblock Digital Video/Audio. McGraw-Hill Education, 2010.

\bibitem{dijkstra}
Edsger~W Dijkstra.
\newblock A note on two problems in connexion with graphs.
\newblock {\em Numerische mathematik}, 1(1):269--271, 1959.

\bibitem{sobel}
Nick Kanopoulos, Nagesh Vasanthavada, and Robert~L Baker.
\newblock Design of an image edge detection filter using the sobel operator.
\newblock {\em IEEE Journal of solid-state circuits}, 23(2):358--367, 1988.

\bibitem{p2020}
{QorIQ P2020 Integrated Processor Reference Manual}.
\newblock {shorturl.at/ltEL0}, 07 2009.

\bibitem{p2020_img}
{QorIQ P2020 Integrated Processor - NXP Page}.
\newblock {shorturl.at/blpMO},
  07 2020.

\end{thebibliography}

\end{document}